\documentstyle[12pt]{article}

\begin{document}

\begin{titlepage}

\begin{flushright}
BRX-TH-381\\
TUTP-94-23\\
hep-th/9510115\\
October 1995
\end{flushright}

\vskip 0.2truecm

\begin{center}
{\large {\bf Black Hole Thermodynamics from Quantum Gravity$^*$}}
\end{center}

\vskip 0.4cm

\begin{center}
{Gilad Lifschytz}
\vskip 0.2cm
{\it Department of Physics,
     Brandeis University,\\
     Waltham, MA 02254, USA.\\
    e-mail: Lifschytz@binach.brandeis.edu }
\vskip 0.3cm
{and}
\vskip 0.3cm
{Miguel Ortiz$^\dagger$}
\vskip 0.2cm
{\it Institute of Cosmology, Department of Physics and
Astronomy,\\
Tufts University, Medford MA 02155, USA\\
and\\
Instituut voor Theoretische Fysika,
Universiteit Utrecht,\\
Postbus 80006,
3508 TA Utrecht, The Netherlands.\\
e-mail: ortiz@fys.ruu.nl}
\end{center}

\vskip 0.8cm

\noindent {\bf Abstract} The semiclassical approximation is studied on
hypersurfaces approaching the union of future null infinity and the event
horizon on a large class of four dimensional black hole backgrounds.
Quantum fluctuations in the background geometry are shown to lead to a
breakdown of the semiclassical approximation in these models. The boundary
of the region where the semiclassical approximation breaks down is used to
define a `stretched horizon'. It is shown that the same effect that brings
about the breakdown in semiclassical evolution associates a temperature and
an entropy to the region behind the stretched horizon, and identifies the
microstates that underlie the thermodynamical properties.  The temperature
defined in this way is equal to that of the black hole and the entropy is
equal to the Bekenstein entropy up to a factor of order one.
\noindent
\vskip 0.8 cm

\begin{flushleft}
\hrulefill

$^*$ {\small This work was supported in part by funds provided by the
U.S. Department of Energy (D.O.E.) under cooperative agreement
DE-FC02-94ER40818, by the National Science Foundation under grants
PHY-9315811 and PHY-9108311, and by the European Community Human Capital
Mobility programme.}

$^\dagger${\small Address from 1/1/96: Blackett Laboratory, Imperial
College, Prince Consort Road, London SW7 2BZ, UK.}
\end{flushleft}

\end{titlepage}

\section{Introduction}

The classical behaviour of black holes contains a number of features that
appear to be in close analogy with the laws of thermodynamics: The
existence of an irreducible mass \cite{chris} proportional to the area of a
black hole event horizon, the fact that in any physical process the area
cannot decrease \cite{hawkinga}, and ultimately the laws of black hole
mechanics \cite{bch}.  On the basis of these classical properties,
Bekenstein \cite{bek} conjectured that the area of the event horizon should
be interpreted as being proportional to an intrinsic entropy of the black
hole, and that there is a generalised second law of thermodynamics that
incorporates black hole entropy. The notion of intrinsic entropy was at
first difficult to reconcile with the view that black holes seem to have
zero temperature, and that they can absorb any amount of entropy thrown in.

The idea that black hole thermodynamics is more than just an analogy was
firmly established by Hawking's demonstration that black holes radiate
\cite{h}. This result was fully compatible with earlier speculations and
fixed the entropy to be one quarter the black hole area.  However,
Hawking's quantum mechanical derivation of black hole radiation arises in a
somewhat different framework to the laws of black hole thermodynamics. Although
it serves as a strong piece of evidence for them, it unfortunately sheds
little light on their microstate origin. Somewhat surprisingly, the problem
of interpreting black hole entropy and of deriving the laws of black
hole thermodynamics from a microstate picture remains an important unsolved
problem.

Important progress in understanding the status of black hole entropy was
made by Gibbons and Hawking \cite{gh}. They showed that the entropy can be
computed via a Euclidean partition function for the gravitational field
with black hole topology. The partition function is evaluated in a saddle
point approximation over gravitational fields with a fixed period $\beta$
in imaginary time. Gibbons and Perry \cite{gp} showed that the temperature
of a black hole can also be derived from the periodicity of the Euclidean
manifold.  Nevertheless, this approach still does not suggest a statistical
interpretation of the first law of black hole thermodynamics or an
interpretation of black hole entropy in terms of microstates.

More recently\footnote{For a recent review on various approaches to black
hole entropy see \cite{bek2}}, there have been a number of attempts to
identify entropies of various kinds with the entropy of a black hole. This
work falls very roughly into two categories.

On the one hand, various attempts have been made to reinterpret the
Gibbons--Hawking calculation as a counting of internal states of the
gravitational field. It has been pointed out by a number of authors that
the non-zero entropy of a black hole has its origins in a surface term at
the horizon of the black hole \cite{hawk,teit}. This suggests
that the internal states of the black hole could be related to degrees of
freedom at the horizon, but no convincing microstate interpretation has yet
to be put forward.

On the other hand, much work has focused on defining the entropy of matter
fields in a black hole background.  This second line of attack has yielded
some interesting results, since the two methods of defining a black hole
entropy using matter fields -- entanglement entropy and statistical
mechanical entropy -- lead to a result proportional to one quarter the area
of the event horizon, provided that an appropriate cut-off is used. The
principal drawback of these approaches is that it is difficult to interpret
such an entropy as the intrinsic entropy of a black hole that appears in
the laws of black hole mechanics. However, they do raise a number of
interesting questions about the interaction between a black hole and
external fields.

In this paper we present some calculations that suggest a microstate
interpretation of black hole thermodynamics. The approach we take falls
somewhere between the two approaches just described\footnote{It is in some
sense similar to a suggestion by Bekenstein \cite{bek2} on the origin of
black hole entropy.}. The origin of thermodynamics is argued to be an
entanglement of the degrees of freedom of matter propagating on a black
hole with those of the black hole \cite{samir,klmo,lo1} (or equivalently
with those of the matter forming the black hole). The entanglement is shown
to lead naturally to a statistical ensemble near the horizon, if one traces
over unobservable fluctuations in the background geometry.

In earlier work, it was shown that on a certain class of hypersurfaces, it
is not possible to compute the state of a quantum field using the
semiclassical approximation of quantum field theory in curved
spacetime. The matter state is highly sensitive to small changes in the
mass of the black hole background, and so a small spread in this mass, due
to uncertainties in the infalling matter distribution that formed the black
hole, leads to an entanglement between the state of the black hole and the
state of the quantum field. This result was derived for a quantum field in
a 2-dimensional black hole background, when the incoming state is in the
Schwarzschild vacuum. The entanglement was shown to involve the infalling
(or left moving) portion of the matter state.

There are a number of points that should be emphasized about these results:
\begin{itemize}

\item The semiclassical approximation breaks down on certain hypersurfaces
in the neighbourhood of the event horizon. These hypersurfaces can be
defined by the condition that they can be extended in such a way as to
capture some of the Hawking radiation at future null
infinity. Alternatively, they can be regarded as constant time slices for a
family of observers whose worldlines remain outside the event horizon.
These hypersurfaces are called S-surfaces.

\item The entanglement arises because propagation on different background
spacetimes with different masses gives rise to matter states that are
different: For a set of backgrounds whose masses are spread within a Planck
mass of some mean value $M_0$, a state evolved to a given spacelike
geometry\footnote{In quantum gravity, which is where this analysis should
take place, a state $\Psi[\phi,h_{ij}]$ is defined as a correlation between
a matter configuration and a spacelike geometry. If for a given $h_{ij}$,
$\Psi_M[\phi,h_{ij}]\ne\Psi_{\bar{M}}[\phi,h_{ij}]$ then there is
entanglement between the $M$ degree of freedom and the matter.} will be
different when evolved on the different backgrounds.

\item Because of the entanglement, there is no canonical matter state on an
S-surface. The matter states for different $M$ are related through a shift
in Kruskal coordinates $x^\pm\to x^\pm+\Delta x^\pm$, where $\Delta^\pm$
are linear functions of the mass fluctuation (a similar shift was derived
under somewhat different assumptions in Refs. \cite{thooft,ver2}).

\item The breakdown or entanglement occurs for the infalling matter sector
close to the horizon.  The shift in Kruskal coordinates $x^+\to x^++\Delta
x^+$ has a large effect on infalling matter that is in (or close to)
the Schwarzschild vacuum. Outgoing matter is in the Kruskal vacuum close to
the horizon and so is not affected by the shift.

\item A different choice of hypersurfaces that foliate the spacetime near
the horizon can give rise to semiclassical evolution, consistent with the
notion that an observer falling through the horizon should not see anything
in conflict with semiclassical physics. There is no problem with using the
semiclassical approximation close to the horizon if one foliates the space
time by hypersurfaces appropriate to an infalling observer, rather than
those appropriate to an outside observer.  This can be expressed as a
quantum gravitational complementarity between observations close to the
event horizon and at future null infinity, or those made by observers at
large relative boosts in the neighbourhood of the horizon.

\item The breakdown in the semiclassical approximation is localized in a
region very close to the event horizon, and its boundary may be regarded as
a stretched horizon. Just
inside the stretched horizon, the breakdown in the semiclassical
approximation becomes dramatic, and the entanglement between matter and
gravity is extremely strong.

\end{itemize}

In this picture, predictions of the physics behind the stretched
horizon cannot be made by observers outside the black hole, using
semiclassical physics.  It was suggested in \cite{lo1} that the
quantum gravitational interactions behind the stretched horizon could
be modeled by assigning some thermodynamic variables to the stretched
horizon. In this paper we discuss the relation between these ideas and
black hole thermodynamics.

The principal results we derive are simple in nature. They are based on the
observation that the results of \cite{klmo} and \cite{lo1} for the
propagation of a Schwarzschild matter state on a black hole, should be
regarded as representing the behaviour of a state that is far from
equilibrium with the black hole. The entanglement computed in
\cite{klmo,lo1} should be regarded as between the matter state and the
internal degrees of freedom of the black hole (represented simply by
different mass eigenstates in a narrow spread around the mean mass of the
black hole). From this point of view, one can perform two simple
calculations.

Firstly, one can look for a state that is not affected by the quantum
gravity effects, and regard such as state as being in equilibrium with the
stretched horizon. It is shown that the Hartle-Hawking vacuum has the
required properties, suggesting that the stretched horizon should be
assigned the temperature associated with this state, which is of course just
the temperature of the black hole.

Secondly, one can count the effective number of internal states of the
black hole by computing the entanglement entropy between the infalling
matter in the Schwarzschild vacuum and the black hole degrees of freedom.
The value of the entropy depends on the choice of S-surface on which it is
computed. The out-of-equilibrium nature of this second calculation is
manifest by construction, and from the fact that in this state, the black
hole radiates. The approximation of a fixed black hole background is not
consistent for evaluating the state on very late time hypersurfaces since
backreaction becomes important. For this reason, one can only obtain an
estimate of the number of internal states of the black hole. This estimate
comes from looking at the entanglement on an S-surface that also catches an
amount of Hawking radiation that is less than (but comparable) to the
original mass of the black hole. Integrating over the small fluctuations of
the black hole mass around some mean mass $M_{0}$, it is found that the
estimate gives an entropy that is equal to the Bekenstein entropy up to a
numerical factor of order unity, and that the result is independent of the
exact details of the surface on which it is evaluated or of number the
matter fields that are used to probe the black hole's internal states.

In the first two sections, we present a short review of the
previous results, extending them to cover a wider class of
models including an arbitrary number of quantum fields in four dimensional
black hole spacetimes. It is shown in Sec. 2 that taking advantage of the
fact that the metric of a black hole spacetime close to its event horizon
takes a generic form, the results of \cite{klmo} and \cite{lo1} on the
embedding of S-surfaces in neighbouring black hole solutions (defined by
slightly different thermodynamic parameters) can be extended to general
static black holes. One exception is the set of extremal black hole
solutions, for which the form of the metric close to the horizon is
qualitatively different. In Sec. 3, the effect of the results of Sec. 2 on
matter propagation is reviewed, and extended to all spherical harmonics
propagating towards a four dimensional black hole. The mathematics of the
inward propagation of matter is closely analogous to the outward
propagation of outgoing matter in  original calculation.

Sections 4 and 5 give details of the calculation of thermodynamic
quantities outlined above. The temperature is easily derived in Sec. 4 from
the results of Sec. 2, since a linear shift in Kruskal coordinates leaves a
thermal state at the black hole temperature unchanged. In Sec. 5 an
estimate is given of the effective number of internal states of the black
hole by computing the entanglement entropy between matter and black hole
states. Tracing over the unobservable degrees of freedom of the black holes
leaves a density matrix describing the infalling matter on a mean
background.  It is also pointed out that the entanglement between matter
and gravity may be a mechanism for the transfer of information from the
black hole to the outgoing matter.

\section{General four Dimensional space time}

We wish to estimate the effects of background fluctuations on the
propagation of matter.  One way to quantify these effects is to compare the
propagation of matter on different classical backgrounds. Because what is
being fluctuated is the spacetime background, more care is required in
making such comparisons than in ordinary quantum mechanics. In quantum
mechanics we can make use of the rigid background coordinate system. In
quantum gravity unambiguous comparisons can also be defined, by comparing
states on hypersurfaces with the same intrinsic geometry.

In this section a sketch is given of how a given spacelike hypersurface,
defined through its intrinsic geometry, is located in a black hole
spacetime as the parameters of the spacetime are changed. The different
locations in different spacetimes can be expressed as a map between
coordinates on the different spacetimes. From this map it is
straightforward to compare quantum states on the given hypersurface.

\subsection{Non-extremal black holes}

For a Schwarzschild black hole, taking $c=1$, the metric
\begin{equation}
ds^2= -\left(1-{2GM\over r}\right)dt^2+\left(1-{2GM\over r}\right)^{-1}dr^2
+r^2\left(d\theta^2+\sin^2\theta d\phi^2\right)
\label{schwarzschild}
\end{equation}
can be approximated close to the event horizon at $r=2M$ by a set of
Rindler co-ordinates, simply by defining $R=r-2GM$, so that
\begin{equation}
ds^2=-\left({2R\kappa}\right)dt^2+\left(2R\kappa\right)^{-1}dR^2+
A_{h}\left(1+{2R\over\sqrt{A_{h}}}\right)d\Omega^2
\label{rindler}
\end{equation}
where $\kappa=1/4GM$ is the surface gravity and $4\pi A_{h}=16\pi G^2M^2$
is the area of the event horizon.  In Minkowski null coordinates the metric
close to the horizon takes the simple form
\begin{equation}
ds^2=-dx^+dx^-+A_{h}\left(1-{\kappa^2x^+x^-\over e\kappa\sqrt{A_{h}}}\right)
 d\Omega^2
\label{nullrindler}
\end{equation}
where $\kappa x^\pm=\pm e^{\pm \kappa t}\sqrt{2e\kappa R}$.
We have neglected a term of order $x^{+}x^{-}$ in the $(t,r)$ part of the
metric. For the type of hypersurfaces we will be
considering, this is a good approximation.

This form of the metric near the horizon is quite general for
static black holes. For example:

\begin{itemize}
\item The metric for the Reissner-N\"ordstrom
spacetime is the same,
 but with $\kappa=\sqrt{(GM)^2-Q^2}/r^2_+$, $r_+=GM+\sqrt{(GM)^2+Q^2}$,
and $A_{h}=r^2_+$.

\item In the CGHS spacetime in 1+1 dimensions, one has $\kappa=\lambda$. In
this case there is no natural definition of the area $4\pi A$, but instead
there is a dilaton field which plays the role of the area, since that is
its origin in a dimensional reduction to the CGHS model.

\item De-Sitter spacetime can be put in a form analogous to (\ref{rindler})
with $\kappa=\sqrt{\Lambda/3}$ but $g_{\Omega\Omega}=
A_{h}\left(1-{2R/\sqrt{A_{h}}}\right)$.
\end{itemize}
\medskip

In what follows we shall consider classical solutions
where a black hole is formed by a pulse of matter, and restrict our
observation to the part of the space time above the pulse (that is for
$\kappa x^+ >1$).

We now turn to the location of spacelike hypersurfaces close to the event
horizon of (\ref{nullrindler}). For simplicity we just look at spherically
symmetric hypersurfaces whose intrinsic geometry is specified by the
function $A(s)\equiv g_{\Omega\Omega}(s)$ where $s$ is the proper distance
along a constant $(\theta,\phi)$ line within the surface from some fixed
point\footnote{The fixed point can be taken to be at infinity \cite{klmo},
although the results we derive are largely independent of this choice. The
choice of fixed point resolves the ambiguity in embedding a hypersurface
with fixed intrinsic geometry in a classical spacetime.}, and $4\pi A$ is
the area of the constant $(t,r)$ 2-surface associated with each point.

We focus on hypersurfaces that are very close to being null surfaces of
constant $x^-$ close to the horizon. Since $x^-=0$ in (\ref{nullrindler})
defines the event horizon, we can look at surfaces of nearly constant $x^-$
as the value of $x^-$ approaches zero. An S-surface is then a hypersurface
for which $\kappa x^-\sim-\delta$, where $\delta$ is very small. This
hypersurface, embedded in a spacetime of mass $M$, has the same intrinsic
geometry as a surface embedded in a spacetime of mass $\bar{M}$ that also
has $\kappa \bar{x}^-\sim-\delta$. However, the identification of different
points within the surface according to their local intrinsic geometry is
given by
\begin{eqnarray}
\bar{\kappa}\bar{x}^+&=&\kappa x^+-\frac{e\kappa(A_{h}-\bar{A}_{h})}
{\delta \sqrt{\bar{A_{h}}}}=\kappa x^+ +
\kappa\Delta x^+ \nonumber
\\
\bar\theta&=&\theta \label{shift}
\\
\bar\phi&=&\phi\nonumber
\end{eqnarray}
so that a given region of the hypersurface is shifted in the $x^+$
direction under the map. In other words a particular region of the
S-surface with a particular local intrinsic geometry located near $x^+_0$
in the mass $M$ spacetime, is located near $\bar{x}^+=x^+_0 +\Delta x^+$ in the
mass $\bar{M}$ spacetime.

For two black hole spacetimes with masses $M$ and $M+\Delta M$ (that are
otherwise identical) the first law of black hole mechanics tells us that
$\bar{A_{h}}= A_{h}(M+\Delta M) \simeq A_{h}(M)+2G\Delta M/\kappa$ (for
black holes not too close to extremality).  Thus the shift in the $x^+$
direction is given by
\begin{equation}
\kappa\left|\Delta x^+\right| =
\left|\frac{2eG\Delta M}{\delta\sqrt{A_{h}}}\right|
\label{sizeofshift}
\end{equation}
This agrees with the exact treatment in the case of the Schwarzschild black
hole \cite{lo1}. This coordinate relationship is valid only for points
above the matter pulse ($\kappa x^+>1$) in both space times.  In order for
an S-surface to have this shift it does not have to be an almost constant
$x^-$ surface throughout. It is enough that it have this property in the
region $1<\kappa x^+<\kappa\Delta x^+$.

Equation (\ref{shift}) shows that for $\delta \ll G\Delta M$ (an S-surface
that runs very close to the black hole), the same hypersurface is embedded
quite differently in the two space times. A generic mass fluctuation
$\Delta M$ should be at the very least of the order of the Planck
mass. $\delta$ is fixed by the choice of S-surface; for an S-surface that
captures some proportion of the Hawking radiation at infinity, $\delta$
must be much smaller than the Planck length and the shift
(\ref{sizeofshift}) is large.

\subsection{The Extremal Black Hole}

The case of extremal black hole solutions is somewhat different because
of the different behaviour of the metric close to the extremal
horizon. Consider for example the extremal Reissner-N\"ordstrom solution
\begin{equation}
ds^2=-\left(1-{GM\over r}\right)^2dt^2+\left(1-{GM\over
r}\right)^{-2}dr^2+r^2d\Omega^2
\label{extremal}
\end{equation}
It is approximated close to the horizon by
\begin{equation}
ds^2=-\left({R\over GM}\right)^2dt^2+\left({GM\over R}\right)^2 dR^2+
A\left(1+{2R\over\sqrt{A}}\right)d\Omega^2
\end{equation}
which is of a different form to Eq. (\ref{rindler}).

It is convenient to define asymptotically flat null co-ordinates
$v,u=t\pm\rho$ where $\rho=r-GM+2GM \ln (r-GM)-G^2M^2/(r-GM)$.
These null coordinates are not
the equivalent of Kruskal coordinates $x^\pm$, which are defined by $v=GM\tan
(x^+/2GM)$, $u=GM\cot(x^-/2GM)$. Close to the event horizon $u\to \infty$ as
$t\to\infty$ and $\rho\to-\infty$.  In terms of $u$ and $v$, the metric
close to the event horizon takes the form
\begin{equation}
ds^2={4(GM)^2dudv\over u^2}+(GM)^2\left(1+{2GM\over
v-u}\right)^2 d\Omega^2
\label{erm}
\end{equation}
On the other hand, in terms of the Kruskal coordinates the metric near the
event horizon is of the form
\begin{equation}
ds^2=\sec^2(x^+/2GM)dx^+ dx^-+(GM)^2\left(1+{2\over
\tan (x^+/2GM)-\cot (x^-/2GM)}\right)d\Omega^2
\end{equation}
We can look at surfaces which close to the horizon are nearly constant
$x^-$ or $u$ lines.  Note that now close to the horizon $x^-\to 0$ but
$u\to\infty$.

Since the topology of an extremal Reissner-N\"ordstrom spacetime is
different to that of a non-extremal hole, it is more natural to consider
fluctuations that do not change the topology, that is that keep $GM=Q$. For
this reason, we examine the identification a spherically symmetric
hypersurface in extremal Reissner-N\"ordstrom solutions with mass $M$ and
$\bar{M}$.

Using (\ref{erm}), and solving for $A(s)=\bar{A}(\bar{s})$ as before, a
surface of constant $x^-$ or $u/GM=\delta$ maps to a surface with the same
$\bar{x}^-$ or $\bar{u}$, but with a shift in the $v$ coordinate of the form
\begin{equation}
\bar{v}=\left(v-{G\Delta M\over
2\delta^2}\right)
\label{ern}
\end{equation}

What is important in this case is not so much the details of the shift but
rather that the linear shift in Kruskal coordinates of the
previous section is replaced by a linear shift in asymptotically flat null
coordinates in the case of an extremal hole. This leads to
qualitatively different black hole thermodynamics in the extremal case as we
shall explain below.

\section{Comparing matter states}

The propagation of the quantum state of a matter field on different
spacetimes can be compared through the same spacelike hypersurfaces, as a
first approximation to the full state functional in quantum gravity. For
free scalar fields in two dimensions, it was shown in \cite{klmo} that the
coordinate relationships defined in the previous section directly define
the result of matter propagation to the given hypersurface. If in a
spacetime of mass $M$ the state is the vacuum with respect to conformally
flat coordinates $x^\mu$, then a fluctuation in the mass of $\Delta M$
transforms the state on the given hypersurface to the vacuum with respect
to $\bar x^\mu(x^\mu)$ where $\bar x^\mu(x^\mu)$ is the coordinate relation
between hypersurfaces in different spacetimes defined in the previous
section.

We shall focus on infalling matter states that start in the Schwarzschild
vacuum at past null infinity. Therefore the relevant coordinate relation is
(\ref{shift}) translated into Schwarzschild coordinates. At ${\cal I}^-$
this relation is approximately the identity, and so there is no effect from
the fluctuations, and the states on any nearby backgrounds are
identical. This remains true on any spacelike hypersurface that is not too
close to the horizon, but is not true for S-surfaces if the shift
(\ref{sizeofshift}) becomes large (the surfaces are too close to the
horizon).

We limit our investigation to four dimensional static spherically symmetric
black hole backgrounds
\begin{equation}
ds^2=g_{tt}(r,t)dt^{2}+g_{rr}(r,t)dr^{2}+r^{2}d\Omega^{2}
\end{equation}
The action for a free scalar field propagating on such background
is
\begin{equation}
S=-\frac{1}{2}\int d^{4}x \sqrt{-g} (\nabla f)^{2}
\end{equation}
and the field can be decomposed as
\begin{equation}
f(\vec{x},t)=\sum_{l,m}
r^{-1}Y_{l,m}(\theta,\phi)f_{l}(r,t).
\end{equation}
where $f_{lm}(r,t)$ solves the equation
\begin{equation}
\left[\frac{d^2}{dr^{*2}}-\frac{d^2}{dt^{2}}+g_{tt}\frac{l(l+1)}{r^{2}}
-\frac{\partial^{2}_{r^*} r}{r}\right]f_{lm}(r,t)=0.
\label{pb}
\end{equation}
Here $r^*$ is the tortoise coordinate.
{}From this one can see that the $f_{0}$ behavior is well approximated by a
free field in two dimensions, while for the $f_{lm}$ with $l>0$ one has to
include a potential term. From now on we shall treat each $f_{lm}$ as an
independent two dimensional matter field.

Since the use of the coordinate relation (\ref{shift}) to compute the
effect of geometry fluctuations on matter states is only valid for free
matter in two dimensions, it can only be applied directly to the $f_{0}$
partial wave.  The $f_{l}$ partial waves with $l>0$ are not free, but we
can use the geometrical optics approximation frequently used in
approximating the Hawking radiation calculation, to get around this
problem.  The propagation of the $l>0$ modes is affected by the potential
in (\ref{pb}).  It is a good approximation to take modes with energy above
the potential barrier to propagate freely and those with energy below the
barrier to not penetrate the barrier at all \cite{dewitt,page}.  Thus when
comparing the two matter states on an S-surface (which always lies behind
the potential barrier) the coordinate relation (\ref{shift}) can be applied
to all modes with enough energy to penetrate the barrier.

Eq. (\ref{shift}) translated to Schwarzschild coordinates reads
\begin{equation}
\bar{v}=\frac{1}{\kappa} \ln \left(e^{\kappa v}-\frac{2e G\Delta M} {\delta
\sqrt{A_{h}}}\right).
\label{asycor}
\end{equation}
As usual the simplest way to compare vacuum states with respect to
different coordinate systems in through the use of
coefficients. Notice that the shift does not mix partial waves. Also, in
the case of the four dimensional black hole we only know the coordinate
relationship (\ref{asycor}) on the part of the S-surface above pulse of
matter that forms the black hole, but this is sufficient for our purposes.

For the relationship (\ref{asycor}) the computation is the same as in
\cite{klmo}, and one finds that for $\Delta M < 0$, in a wave packet basis,
the Bogoliubov coefficients are thermal (with temperature
$T={\hbar\kappa}/{2\pi}$)\footnote{One should not take the thermal
character to mean that one state is pure and the other is mixed. It is just
an indication that we are computing the overlap of states on the region of
the S-surface above the pulse of matter and are ignoring correlations with
the part of the S-surface that we ignore.}, in a region of $\kappa v$ of
size $\ln(\kappa \Delta x^+)$.  That is
\begin{equation}
\beta^{*}_{jn,w'} \approx -e^{-\pi w_{j}/\kappa} \alpha_{jn,w'}
\label{bog1}
\end{equation}
where $(j,n)$ parameterizes the wave packets in $v$ ($w_{j}=ja$ and the
wave packet is centered around $v=2\pi n/a$ with spatial width $\sim
a^{-1}$) and $w'$ is a continuous parameter labeling the modes in
$\bar{v}$.  In the case $\Delta M > 0$ the result is the same, but the
roles of $v$ and $\bar{v}$ are interchanged.

Using the relationship
\begin{equation}
\int_{0}^{\infty} dw'' \alpha_{jn,w''} \alpha^{*}_{j'n',w''} - \beta_{jn,w''}
\beta^{*}_{j'n',w''}= \delta_{jj'}\delta_{nn'},
\end{equation}
one gets
\begin{equation}
(\alpha \alpha ^{\dagger})_{jn,j'n'} \approx
\frac{\delta_{jj'}\delta_{nn'}} {1-e^{-2\pi w_{j}/\kappa}},
\end{equation}
from which the inner product between the two states on the S-surface given
by
\begin{equation}
\ln\left|\left\langle 0_{in}, l, M\Bigl|\bar{M}, l,
0_{in}\right\rangle \right|^{2}= ({\rm det}(\alpha \alpha
^{\dagger}))^{-\frac{1}{2}}
\end{equation}
Notice that this quantity does not need to be regularised as the
divergences are related to the imaginary part of the overlap
\cite{dewitt}. Although this expression only gives the overlap in the
region above the matter pulse that formed the hole, it is a good
approximation to the total overlap between the
states.

Following \cite{klmo}, the inner product between two matter states for the
field $f_{l}$ as compared on an S-surface is given by
\begin{equation}
\ln\left|\left\langle 0_{in}, l, M\Bigl|\bar{M}, l,
0_{in}\right\rangle \right|^{2}=\frac{1}{2}n_{max} \sum_{j} \Gamma_{jl} \ln
(1-e^{-2\pi w_{j}/\kappa})
\end{equation}
where $\Gamma_{jl}=\theta(bw_{j}-l)$ (in the case of Schwarzschild black
hole $b=\sqrt{27}GM$ \cite{page}), and where
\begin{equation}
n_{max}=\frac{\ln\left(\kappa\Delta x^+\right)}{2\pi(\kappa/a)}.
\end{equation}
This can be evaluated approximately to give
\begin{equation}
\left|\left\langle 0_{in}, l, M\Bigl|\bar{M}, l, 0_{in}\right\rangle
\right|^2= \exp \left( -e^{-(2\pi l/\kappa b)}\frac{\ln(\kappa \Delta
x^+)}{8\pi^{2}}\right)
\label{inerprol}
\end{equation}
for $l \gg 0$, which approaches $1$ for a fixed hypersurface as $l$
increases.  For $l=0$ \cite{klmo},
\begin{equation}
\left|\left\langle 0_{in}, l=0, M\Bigl|\bar{M}, l=0,
0_{in}\right\rangle \right|^2=\exp\left(-\frac{\ln(\kappa \Delta
x^+)}{48}\right)
\end{equation}

One can define approximate orthogonality by the condition that the inner
product should be less than some number $\gamma\ll 1$, and ask for a given
S-surface how large the fluctuation $\Delta M$ should be for two states
propagated on spacetimes differing in mass by $\Delta M$ to be
approximately orthogonal. Recall that an S-surface is defined by $\delta$,
where $\kappa x^-\sim -\delta$, which is related to $\Delta x^+$ by
(\ref{sizeofshift}).  For the inner product to be of order $\gamma$,
$\Delta M$ needs to satisfy
\begin{equation}
\Delta M ={\delta \sqrt{A_H} \gamma^{-8\pi^{2}e^{2\pi l/\kappa
b}}\over G}
\end{equation}
Supposing that the fluctuation in $M$, $\Delta M$, is of order the Planck
mass, the number of different states in the $l$th partial wave on the
S-surface is
\begin{equation}
{\cal N}_{lm} \approx {1\over\delta}
\left({\hbar G\over A_H}\right)^{1\over 2}
\gamma^{8\pi^{2}e^{2\pi l/\kappa b}}
\label{numst1}
\end{equation}
For the case of the $f_{0}$ field the exponent of $\gamma$ is $48$.  For an
S-surface catching some proportion of the Hawking radiation $\delta$ is a
very small dimensionless number. Thus for the lower partial waves, the
number of states in (\ref{numst1}) is very large.

The validity of the semiclassical approximation depends on assuming that
the Hilbert space structure of the matter fields on a given hypersurface
does not depend on the small fluctuations of the gravity sector. That is,
there should be a unique matter state for any given hypersurface. Equation
(\ref{numst}) shows that this is not the case on S-surfaces. What is worse,
there are many different possible matter states that can be
defined on the S-surface and that are compatible with a black hole of mass
$M$ defined up to Planck scale fluctuations. The two dimensional version of
this result is discussed in Refs. \cite{klmo,lo1}. There are a number of
points related to this observation that can be added to those made in the
introduction.

\begin{itemize}

\item Since there is no uniquely defined state on an S-surface, the
semiclassical approximation is breaking down \cite{klmo}. However, it is
possible for two states to be orthogonal but to respond in a very similar
way to all interesting physical operators. In \cite{lo1} it was shown that
the states considered above have very different energies when regulated in
the same way. This in itself shows that the two states are not only
orthogonal but respond differently to physical operators.

\item The regions of large energy found in \cite{lo1} coincide with the
part of the S-surface where the Bogoliubov coefficients appear
thermal. The boundary of this  region  can be seen from
equation (\ref{shift}) to be the timelike
hypersurface with area $4\pi A= 4\pi A_{h}(GM+G\Delta M_{max})$.

\item The boundary surface can be thought of as a stretched horizon. Behind
the stretched horizon, quantum gravity becomes important. Without a theory
of quantum gravity the effect of the region behind the stretched horizon
on the physics at ${\cal I}^+$ must be described by an effective theory.

\item If one integrates over the small fluctuations of the mass of the
black hole around the mean spacetime with mass $M_{0}$, then the resulting
matter state is a density matrix which is pure at ${\cal I}^{-}$ and
becomes a statistical ensemble of many different states on the portion of
the S-surface behind the stretched horizon.

\end{itemize}

It is tempting to speculate that this effective theory behind the stretched
horizon has a thermodynamical character.

\section{Temperature}

Given a box with some thermal properties one can measure its temperature,
pressure, chemical potential, etc., by bringing it into contact with
another system. If the temperature of the system, say, is known, and the
system is unchanged after coming into contact with the box, one would say
that the two systems are in equilibrium and that they have the same
temperature. Notice that by this procedure one cannot directly measure any
extensive properties like energy, entropy, etc. For these a different
approach must be taken.

We have seen in the examples given in Sec. 2.1 that the coordinate
relationship that results when identifying the same hypersurface (near the
horizon) in two different space times is a linear shift in the Kruskal
coordinates. For a state that starts as the vacuum with respect to $\kappa
v \sim \ln \kappa x^{+}$ at ${\cal I}^{-}$, this results in the
breakdown of the semiclassical approximation on S-surfaces, and the
occurrence of a statistical ensemble of states near the horizon. An outside
observer sees a field interacting with a stretched horizon in a way that we
cannot determine without quantum gravity.

If the incoming matter state is in the Kruskal vacuum the linear shift has
{\it no} effect, thus an outside observer sees the Kruskal vacuum
unaffected by the stretched horizon. In thermodynamical language the
stretched horizon and the Kruskal vacuum can be said to be in
equilibrium. As is well known the Kruskal vacuum (restricted to the part
outside the horizon) corresponds to a thermal matter state that has local
temperature $T_{local}=(g_{tt})^{-1/2} T_{BH}$ \cite{HarHawk}.  It is
therefore natural to regard the stretched horizon as a constant temperature
hypersurface, with the same temperature $T_{local}$ as the local
temperature of the Kruskal vacuum.

It is interesting to note that since the region of future null infinity is
to the causal future of the stretched horizon, then within this
picture, quantum gravity effects
should somehow affect the outgoing Hawking radiation. Nevertheless, there
are two pieces of evidence that indicate that the black hole does radiate
at a temperature $T$ (which is not to say that the radiation is necessarily
exactly thermal). The first is that the stretched horizon seems to be at
a temperature consistent with Hawking radiation. The second is that so far
the only effect we have computed on the outgoing matter is a shift that
does not change the Hawking state.

Finally, note that according to this criterion, the equilibrium state for
an extremal black hole is the Schwarzschild vacuum which is at zero
temperature. If one imagines global fluctuations in De-Sitter space (say
due to a fluctuating cosmological constant, or the appearance
of a small De-Sitter black hole), the state that is in
equilibrium with the cosmological horizon is then the De-Sitter invariant
vacuum as might be expected.

\section{Entropy}

Let us now turn to the relationship between black hole entropy and the
entanglement between the degrees of freedom of the black hole and of matter
fields propagating on the hole. In the preceding calculations, the
degrees of freedom of the hole have been associated with the continuous
parameter $M$, and the state for the black hole has been assumed to be a
superposition of eigenstates of $M$ with a small spread $\Delta M_{max}$
around some mean value $M_0$.

In section 3 we have estimated the degree of entanglement by computing the
number of approximately orthogonal matter states which are produced on an
S-surface $\Sigma$ by propagation on black hole backgrounds with masses in
the range $(M-\Delta M_{max},M+\Delta M_{max})$. On general grounds $\Delta
M_{max}$ should be taken to be at least of the order of the Planck mass.

{}From equation (\ref{numst1}) the log of the number of approximately
orthogonal states for a field $f_{lm}$ on a particular S-surface is
given by
\begin{equation}
S^{Ent}_{lm} \equiv \ln {\cal N}_{lm} = \ln\left(\kappa \Delta
x^+_{max}\right)+8\pi^{2}e^{2\pi l/\kappa b} \ln\gamma
\label{numst}
\end{equation}
where
\begin{equation}
\kappa \Delta x^+_{max}= {2e\over\delta}
\left({\hbar G\over A_H}\right)^{1\over 2}.
\end{equation}
This quantity approximates the entanglement entropy obtained by tracing
over the degrees of freedom of the black hole and computing the entropy of
the resulting density matrix for each partial wave.

The entanglement entropy (\ref{numst}) in each partial wave depends on the
choice of S-surface through the parameter $\delta$. It is clear that as
$\delta\to 0$ (as the S-surface comes arbitrarily close to the horizon),
the entropy diverges. However since the S-surfaces collect Hawking
radiation from the black hole at future null infinity, this leads to a
backreaction on the black hole background. For this reason, in a non
evaporating black hole background, it does not make sense to look at
S-surfaces which capture an amount of Hawking radiation greater than the
mass of the black hole. This imposes a lower bound on the value of
$\delta$.  We shall take the entropy for the smallest $\delta$ to be an
indication of the maximum number of internal states of the black hole.

\subsection{Entropy in 2 dimensions}

As a first example consider only the $l=0$ partial wave, which is equivalent
to looking at a 2-dimensional field theory. In that case (\ref{numst}) is
equal to $\ln(\kappa\Delta x^+_{max})$, which is interpreted as the quantum
gravitational entropy for the $l=0$ partial wave, and depends on $\delta$.
Characterizing an S-surface by the amount of energy $E(\delta)$ it captures
in Hawking radiation, we can define $S^{Ent}_{l=0}(E(\delta))$.  For a
2-dimensional field theory \cite{cghs},
\begin{equation}
E={u_{tot}}\frac{\hbar\kappa ^{2}}{48 \pi}
\end{equation}
where $u_{tot}$ is length of retarded time (at ${\cal I}^{+}$) for which
the S-surface catches Hawking radiation at the rate
$\hbar\kappa^2/48\pi$. For S-surfaces $\kappa u_{tot} \approx \ln(\kappa
\Delta x^+_{max})$ so that and S-surface with
\begin{equation}
\delta\sim 2\left({\hbar G\over A_H}\right)^{1\over 2}
e^{-{48\pi E\over\hbar\kappa}}
\end{equation}
captures an energy $E$ in Hawking radiation. It follows that the
entanglement entropy and the energy captured at infinity are related by
\begin{equation}
S^{Ent}_{l=0}(E(\delta))={1\over 2}\ln\left[{4e^2\hbar G\over
\delta^2 A_H}\right]= \frac{48\pi E}{\hbar\kappa}
\label{entropy1}
\end{equation}
Note that $S$ scales like $1/\hbar$ as expected.

\subsection{$N$ matter fields and the large $N$ approximation}

For $N$ identical matter fields in 2 dimensions, all contributing the same
amount of energy on the S-surface, the number of approximately orthogonal
states is given by
\begin{equation}
{\cal N}_{tot}={\cal N}_{1}*{\cal N}_{2}* \cdot *{\cal
N}_{N}={\cal N}_1^N
\end{equation}
where by assumption the ${\cal N}_i$ are all equal. Because there are more
matter fields present, the energy flux at ${\cal I}^+$ scales as $N$, so
that an S-surface with
\begin{equation}
\delta\sim 2\left({\hbar G\over A_H}\right)^{1\over 2} e^{-{48\pi E\over
N\hbar\kappa}}
\end{equation}
captures an energy $E$ in Hawking
radiation. Thus the relation between the entanglement entropy and the
energy captured at infinity is
\begin{equation}
S^{Ent}_{N}(E(\delta))={N\over 2}\ln\left[{4e^2\hbar G\over
\delta^2 A_H}\right]= \frac{48\pi E}{\hbar\kappa}
\label{entropy2}
\end{equation}

This result is a first indication that the entanglement entropy for
an S-surface that captures an amount of Hawking radiation comparable with
the mass of the black hole (smallest value of $\delta$) is independent of
the details of the matter fields that are used to compute it. This is a
basic requirement if the entanglement entropy is to be related to an
intrinsic property of the black hole.

This result may appear to contradict the expectation that the semiclassical
approximation is exact in the limit of large $N$. However, in order to take
the large $N$ limit, it is important to keep track of factors of $\hbar$,
since $\hbar N$ is required to remain fixed at large $N$ \cite{gidd}.  For
the $i$th matter field, the number of approximately orthogonal matter
states on an S-surface catching a total energy $E$ (in all fields) is
\begin{equation}
{\cal N}_i \approx \sqrt{\hbar} \exp\left(\frac{48\pi E}
{\hbar \kappa N}\right)
\label{largen}
\end{equation}
Equation (\ref{largen}) is valid if the number of states is large.
Really one should add a $1$ to equation (\ref{largen}) to count the
semiclassical state (for which $\Delta M =0$). If any
${\cal N}_i$ becomes less than 1, it means that there is only one state,
the semiclassical state defined on any spacelike hypersurface. If ${\cal
N}_i < 1$ for all matter fields, then there is no gravitational entropy,
and the semiclassical approximation is valid. Now, as $\hbar \rightarrow 0$
with $\hbar N$ held fixed ${\cal N}_i \rightarrow 0$.  As soon as ${\cal
N}_i$ drops below 1, the number of states for any one of the matter fields
is then just one, the semiclassical state. Hence in this limit the we do
not expect to see deviations from the semiclassical approximation, as
expected.

\subsection{Entropy in 4 dimensions}

For a 4-dimensional black hole, one can repeat the previous calculation,
but taking into account the contribution of the higher partial waves to the
entanglement entropy and to the Hawking radiation. The energy collected at
${\cal I}^+$ is given by
\begin{equation}
E_{lm}={u_{tot}\over 2\pi}\int\frac{dw \;\hbar
w\Gamma_{lm}(\omega)}{e^{\hbar w/T}-1} \equiv
{u_{tot}\tilde{\Gamma}_{lm}\pi T^{2}\over 12\hbar}
\label{enerlm}
\end{equation}
where $u_{tot}$ represents the total retarded time for which the S-surface
catches Hawking radiation at the rate $\pi \tilde{\Gamma}_{l,m}T^2/12\hbar$
and $\tilde{\Gamma}_0=1$.  The total energy for all modes is thus
\begin{equation}
E=\left({u_{tot}\pi T^2\over 12\hbar}\right)\sum_{lm}
\tilde{\Gamma}_{lm}={u_{tot}\pi T^2\over 12\hbar}\sigma_{tot}
\end{equation}
where ${u_{tot}\pi T^2/ 12\hbar}$ is just the energy contributed by the
$l=0$ partial wave. For the Schwarzschild black hole, for example,
$\Gamma_{lm}=\Theta(\sqrt{27}wGM-l)$, so that $\sigma_{tot}\sim
AT^{2}/\hbar^2$ which gives the usual $3+1$ dimensional black body result
that the hole radiates at a rate proportional to $AT^4/\hbar^3$.

To compute the entropy as a function the energy radiated at infinity, we
fix the energy to be some $E_{0}$. This can be taken to be either an
infinitesimal amount $dM$ or up to the entire mass of the black hole (for
uncharged holes). Each partial wave contributes energy
(\ref{enerlm}). Thus we are looking for an S-surface with
\begin{equation}
\ln(\kappa\Delta x^+_{max})=\kappa u_{tot}=\frac{48\pi
E_{0}}{\sigma_{tot}\hbar\kappa}
\end{equation}
The total entanglement
entropy for such a surface can be estimated as follows. In each partial
wave there are $\exp (S^{Ent}_{l,m}(\Sigma))$ approximately orthogonal
states according to equation (\ref{numst}). There are an infinite number of
partial waves, but the potential barrier (\ref{pb}) reflects an increasing
proportion of quanta in the higher $l$ partial waves. Since the potential
barrier for the infalling matter is the same as that for the outgoing
Hawking radiation, the total entropy can be estimated to be
(in the microcanonical ensemble)
\begin{equation}
S^{Ent}_{tot}(\Sigma)=\sum_{lm}\tilde{\Gamma}_{lm}S^{Ent}_{lm}(\Sigma)
\end{equation}
and using (\ref{numst}), the leading term is given by
\begin{equation}
S^{Ent}_{tot}(E_{0})=\frac{48\pi E_{0}}{\hbar\kappa}.
\label{fentro}
\end{equation}
which is again the same as (\ref{entropy1}).

In the case of Schwarzschild black hole, an estimate of the total number of
internal states is given by taking $E_0=M$. In that case,
\begin{equation}
S\sim A/\hbar
\end{equation}
with a constant of proportionality which is $12$.

In the case of the charged black hole, it is more difficult to choose a
reasonable value of $E_0$ to obtain an estimate for the number of internal
states in the case where the radiation is through uncharged
fields. Empirically we find that taking $E_{0} \sim (M-Q\Phi)$ gives a
result that is in agreement with the entropy of a charged black hole.

\subsection{Entropy of extremal black holes}

The analysis of this section cannot be applied to an extremal black
hole. In the case of the extremal Reissner-N\"ordstrom solution, equation
(\ref{ern}) describes the effect of fluctuations of the geometry that
preserve the topology of the extremal hole. This effect consists of a large
shift along S-surfaces, but that is linear in the coordinate $v$ that
defines the natural incoming vacuum from ${\cal I}^{-}$. Thus the natural
incoming state does not become entangled with the gravitational degrees of
freedom, implying that the extremal hole has no effective degrees of
freedom. This result appears to be compatible with recent results arguing
that extremal black holes have zero entropy from the standard viewpoint
\cite{teit}.  Recall that in the previous section it was shown that the
stretched horizon is at zero temperature.

\section{Conclusions}

We have shown that the breakdown of the semiclassical approximation, for
surfaces that capture some Hawking radiation and stay outside the black
hole event horizon (an S-surface), is a phenomenon common to both two and
four dimensional black holes.  Further, we have suggested that the same
mechanism that is responsible for the breakdown in the semiclassical
approximation can account for the thermodynamical properties of black
holes. The quantum gravitational degrees of freedom of the black hole
interact with matter fields in such a way that the black hole appears to
have a large number of different internal states. The log of the number of
internal states is equal to the black hole entropy up to a numerical factor
of order unity. These states can perhaps also be thought as the different
matter states that can form a black hole with mass between $M$ and
$M+\Delta M_{max}$.

These results suggest that a microstate interpretation of black hole
entropy is a simple consequence of regarding a black hole of mass $M$ as
being in a superposition of mass eigenstates with a Planck sized spread
around $M$ that is certainly unobservable. The interaction between the
microstates of the gravitational field and external matter fields can only
be crudely approximated without a complete theory of quantum
gravity. However, it is plausible that an effective theory of black hole
evaporation can be formulated in terms of a stretched horizon separating
the semiclassical region from the region of strong quantum gravitational
interactions. The fact that the Kruskal vacuum is unaffected by quantum
gravitational fluctuations suggests that this stretched horizon has an
effective temperature equal to that of the black hole. In this picture
(which is very similar to the one advocated in \cite{stu})
the stretched horizon is in causal contact with the outside matter fields,
and is responsible for the thermodynamical properties of the black hole.
It is encouraging
that the temperature and number of internal states of the black hole can
both be derived from a single calculation.

To make the preceding statements more concrete, it would be necessary to
find a relation between them and what is currently understood about black
hole entropy. At present this is not understood. For example once the
microstate origin of black hole entropy is completely understood, it should
be possible to derive the laws of black hole mechanics, and also to
relate the entropy to the Euclidean calculations of black hole entropy.  On
this second question, it is interesting to note that the large entanglement
found close to the horizon is a direct consequence of the fact that the
horizon is a lightlike constant $r$ surface (a degenerate two surface), the
same feature that leads to non-zero entropy from the Euclidean partition
function.

Finally, the entanglement between quantum gravitational degrees of freedom
and matter fields on an S-surface can be seen to be equivalent to an
entanglement between the matter state that forms the black hole and the
matter state on an S-surface. The relation between the entropy of
entanglement and the amount of energy collected on the S-surface in the form
of Hawking radiation, and the possible origin of the black hole
entropy, is a hint that in a more precise calculation,
information may be being transferred from the black hole to the matter
fields.

\section*{Acknowledgements}

G.L. would like to thank Adi Stern for many helpful discussions and the Center
for Theoretical Physics, MIT for financial support during the first stages
of this work.

\end{document}